\documentclass[letterpaper,twocolumn,10pt]{article}
\usepackage{usenix2019_v3}

\usepackage{tikz}
\usepackage{amsmath}
\usepackage{amsfonts}
\usepackage{xcolor,colortbl}
\usepackage{url}
\usepackage{svg}
\usepackage{paralist}
\usepackage[font=small]{caption}
\usepackage[font=footnotesize]{subfig}
\usepackage[inline]{enumitem}

\usepackage[noorphans,vskip=0.5ex]{quoting}

\usepackage{filecontents} 

\usepackage{diagbox}

\newif\ifshowcomments
\showcommentsfalse
 
\ifshowcomments
    \usepackage[textsize=small,textwidth=0.60in]{todonotes}
    \newcommand{\David}[2][]{\todo[color=blue!30,#1]{David: #2}}
    \newcommand{\Wendy}[2][]{\todo[color=orange!30,#1]{Wendy: #2}}

\else
    \newcommand{\David}[2][]{\ignorespaces}
    \newcommand{\Wendy}[2][]{\ignorespaces}

    \newcommand{\todo}[1]{\ignorespaces}

\fi

\newcommand{\ourtool}{Calpric}

\begin{document}

\date{}

\title{\Large \bf \ourtool{}: Inclusive and Fine-grained Labeling of Privacy Policies with Crowdsourcing and Active Learning}

\author{
{\rm Wenjun Qiu\hspace{0.5in}David Lie\hspace{0.5in}Lisa Austin}\\
University of Toronto
} %

\maketitle 

\begin{abstract}
A significant challenge to training accurate deep learning models on privacy policies is the cost and difficulty of obtaining a large and comprehensive set of training data.  To address these challenges, we present \ourtool{}, which combines automatic text selection and segmentation, active learning and the use of crowdsourced annotators to generate a large, balanced training set for Android privacy policies at low cost.  Automated text selection and segmentation simplifies the labeling task, enabling untrained annotators from crowdsourcing platforms, like Amazon's Mechanical Turk, to be competitive with trained annotators, such as law students, and also reduces inter-annotator agreement, which decreases labeling cost. Having reliable labels for training enables the use of active learning, which uses fewer training samples to efficiently cover the input space, further reducing cost and improving class and data category balance in the data set.

The combination of these techniques allows \ourtool{} to produce models that are accurate over a wider range of data categories, and provide more detailed, fine-grained labels than previous work.   Our crowdsourcing process enables \ourtool{} to attain reliable labeled data at a cost of roughly \$0.92-\$1.71 per labeled text segment.  \ourtool{}'s training process also generates a labeled data set of 16K privacy policy text segments across 9 data categories with balanced assertions and denials. 
\end{abstract}

\section{Introduction}

Privacy policies are legal documents that disclose how a party collects, uses, and shares personally identifiable information (PII). Privacy legislation, such as the California Online Privacy Protection Act (CalOPPA) and the General Data Protection Regulation (GDPR), require digital services to use privacy policies to obtain consent for the collection and use of PII. In addition to consent, privacy policies also serve as a mechanism for transparency and accountability to ensure that digital services comply with privacy legislation~\cite{lie:utlj2021:automated_accountability}. With the massive growth of privacy policies, there has been significant interest in building tools to automate the analysis of privacy policies for compliance~\cite{zimmeck_automated_2017, zimmeck2019}, as well as to help end users understand them~\cite{harkous2018}.

However, previous attempts have been limited in inclusiveness and granularity\footnote{We use inclusiveness to denote accurate classification for minority data categories and fine/coarse to denote the number of possible labels a model can predict at its output.}. For example, \cite{zimmeck2019} and \cite{zimmeck_automated_2017} are limited to only the most common categories of private data---contacts, device identifiers and location---but do not include less common data categories such as private health, financial and demographic information. Similarly, \cite{harkous2018} trains hierarchical models that classify text segments by various privacy-related properties, such as the type of \textit{data action} (i.e. collection, storage) and the \textit{data category} (location, health). However, their models cannot distinguish fine-grained attributes, such as an explicit denial of a data action from the lack of a disclosure of a data action. In addition, they also cannot guarantee uniform performance on rare combinations (i.e.``We do not store your health information'', since explicit denials and storage are both rarer classes). 
PolicyLint~\cite{policyLint} and PoliCheck~\cite{PoliCheck} use a rules-based ontology to organize information types as a hierarchy, but also do not explicitly address minority data categories. These shortcomings exist because 
\begin{enumerate*}[label=\arabic*)]
\item the training set in previous approaches is not sufficiently large to cover less common cases, and 
\item the imbalanced distribution of cases in privacy policies means that there will be insufficient training samples to provide good performance at fine-granularity.
\end{enumerate*}

\begin{figure}
	\centering
	\includegraphics[scale=0.47]{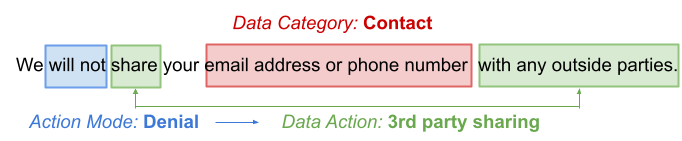}\\[-1ex]
	\caption{A fully labeled privacy policy segment with data category (contact information), data action (third party sharing), and its corresponding action mode (does not share).}
	\label{fig:category_example}
\end{figure}

In this paper, we address both challenges with a new training method for privacy policy classifiers, which we call \textit{\ourtool{}} (\underline{C}rowdsourcing \underline{A}ctive \underline{L}earning \underline{PRI}vacy Policy \underline{C}lassifier). \ourtool{}'s training approach simultaneously gathers data and trains on it, synergistically using crowdsourced annotators and active learning to produce a larger and distributionally balanced dataset. \ourtool{} overcomes a number of challenges that arise from crowdsourcing, such as untrained annotators and unreliable annotations, through a data category-specific text segmentation algorithm.  The construction of previous data sets, such as OPP-115 and APP-350~\cite{wilson2016, zimmeck2019}, required the annotators to both select the relevant text segments in the privacy policy and label them, a complex annotation task for crowdsourced workers. \ourtool{} automatically extracts segments comparable to the ones selected by trained annotators by determining whether a segment is relevant to the label being sought, identifying the correct segment boundary and capturing the necessary local context. However, simply acquiring more training samples would not guarantee that sufficient samples of rare categories are found. To overcome distributional imbalance, \ourtool{} uses active learning, which uses a partially trained model to select unlabeled samples that are likely to contain information that the model currently lacks. The use of active learning itself is not without pitfalls---because active learning strategically selects samples for which it currently has little information, there tends to be little redundancy in the training samples it selects, and so an incorrect label provided by an annotator can have a large impact on model accuracy. Unfortunately, crowdsourced annotators do not provide correct labels consistently.  \ourtool{}'s segmentation algorithm overcomes this challenge by increasing inter-annotator agreement and annotation accuracy, thus providing benefits to both the crowdsourcing and active learning components.

Building on these innovations, we use \ourtool{} to build the \ourtool{} Privacy Policy Corpus (CPPS). Training on CPPS, \ourtool{} produces models that can provide accuracy over a greater breadth of data categories and with finer-grained labels than previous work.  \ourtool{} achieves an overall classification accuracy of $93.0\%$ on the sentence/segment level, and $92.2\%$ on the document level, which exceeds that of prior models trained on smaller data sets generated by trained annotators~\cite{wilson2016,zimmeck2019,harkous2018}. 
Simultaneously, \ourtool{}'s training methods enable the use of crowdsourced annotators labeled text segment for \$0.92-\$1.71 at an annotation quality similar to previous studies that used used more expensive trained annotators~\cite{zimmeck_automated_2017,zimmeck2019}. This enables \ourtool{} to generate a training set set of 16856 labeled text segments, which is 3$\times$ larger than the largest previous human-annotated data set and contains significantly more information on minority data categories and classes.

\vspace{4pt}
\noindent In summary, this paper makes the following contributions: 

\begin{enumerate}
	\item We present \ourtool{}, which synergistically combines crowdsourcing and active learning to produce \ourtool{} Privacy Policy Corpus (CPPS), which contains 16856 labeled text segments extracted from 52K Android privacy policies covering  9 data categories and 3 data actions.  CPPS is balanced across data categories and action modes. As far as we are aware, CPPS is the largest corpus of privacy policy text segments to date. The \ourtool{} project is openly available \url{https://github.com/dlgroupuoft/Calpric}.  
	\item  While creating CPPS, \ourtool{} trains models that provide more inclusive and finer-grained classification performance on privacy policies. \ourtool{}'s models cover 9 data categories and can differentiate denials (i.e. explicit denial of a data action) from the absence of disclosure about a data action.
	\item We perform measurements and find that \ourtool{}'s automated text segmentation increases the number of usable labels produced by 65\%, enables the use of crowdsourced annotators, which reduces labeling costs by 9$\times$, and results in a model that is more accurate than prior work, especially on minority data categories and classes. 
    \item We conduct a study on rare data categories and actions in 1,800 Android app privacy policies. We find that some categories are not as rare as previously believed, and that with the exception of the most popular applications, the number of explicit denials and use of controllable data actions increases with application popularity.
\end{enumerate}

\section{Challenges}
In this section we better characterize the challenges of inclusive and fine-grained privacy policy classification and the availability of training data, which covers data imbalance, labeling cost, and segmentation challenges.  

\subsection{Inclusive and Fine-grained Classification} \label{challenge:breadth_fine_classification}

While previous work has demonstrated that it is possible to train models that achieve high classification accuracy on privacy policy text, that classification has been limited in inclusiveness and granularity. We summarize the limitations of previous approaches as compared to \ourtool{} in Table \ref{tab:tools_comparison}. Limitations in breadth mean that model predication accuracy may suffer for minority (rare) classes, as machine trained models tend to optimize for accuracy in majority (common) classes. \cite{zimmeck2019} and \cite{zimmeck_automated_2017} only cover 3 common categories and do not cover rare data categories such as health, financial, demographic, social media, personal identifier, or survey data. \cite{harkous2018} covers 11\footnote{Polisis covers non-mobile privacy policies, we only train \ourtool{} on categories that are applicable to mobile privacy policies.} rare categories, but has low accuracy in them due to the limited access to training data in those categories. 

To alleviate some of these issues \cite{harkous2018} combines different data actions when training models to predict certain private attributes. For example, a single ``information type'' attribute model is trained across all data actions, combining actions such as 1st party (common) and 3rd party (rare) collection. As a result, they can only report prediction accuracy overall, and cannot guarantee accurate prediction for 3rd party collection.  

\begin{table}[]
\footnotesize
\begin{tabular}{|p{.83in}|p{1.35in}|p{.65in}|}
\hline
           & Granularity of labels                              & Categories              \\ 
\hline
Polisis~\cite{harkous2018}   & Relevant / $\neg$Relevant (2)                      & Common (3) + Rare (11)  \\ 
\hline
MAPS~\cite{zimmeck2019}/\cite{zimmeck_automated_2017}       & Assertion / Denial (2)                            & Common (3)              \\ 
\hline
PolicyLint~\cite{policyLint} & Assertion / Denial (2)                            & All\footnote{}                     \\ 
\hline
Calpric    & Assertion / Denial / Non Mentioned / Choice / Ambiguous (5) & Common (3) + Rare (6)   \\
\hline
\end{tabular}\\[-1ex]
\caption{Comparison of automated privacy policy analysis tools}
\label{tab:tools_comparison}
\end{table}

\footnotetext{Technically PolicyLint and PoliCheck are able to handle all categories, but they rely on name-entity recognition, which does not take into account context like our language-model based extraction.}

Insufficient data and biased distribution also lead to coarser-grained categories, which provide less information to the end user. Previous machine learning approaches generally cannot identify phrases with explicit denials (``We do not collect your device identifier information'')~\cite{zimmeck_automated_2017,harkous2018}. While \cite{zimmeck2019} and PolicyLint~\cite{policyLint} can identify explicit denials, they do so using explicit rules to either generate synthetic denial samples, or check for negative modifiers in a dependency parse tree. In addition, none can extract indications of choice or opt-out options (``You may opt-out of this collection''), or ambiguity (``We may not collect your contact information''). In contrast, \ourtool{} is able to differentiate between statements that (i) assert collection; (ii) explicitly deny collection; (iii) provide an opt-out option; or (iv) are ambiguous.

\ourtool{}'s training approach gathers data at lower cost through crowdsourcing and produces a dataset with a balanced distribution across target categories via crowdsourcing and active learning. However, the use of these two approaches themselves create new challenges, which we detail below.

\subsection{Cost and Imbalance in Training Data}

We detail the challenges of distributional imbalance in training data and the cost of acquiring labels for training.

\subsubsection{Class Imbalance} \label{challenge:imbalance}

OPP-115~\cite{zimmeck_automated_2017} and APP-350~\cite{zimmeck2019} are so far the most used privacy policy data sets for machine learning-based models, yet they both suffer from data imbalance. That is, the distribution of \textit{data categories} collected by digital applications is not uniform, and as a result, privacy policies do not uniformly cover various categories of data. Imbalance exists across both data categories and positive/negative classes. For example, there are many more privacy policy text segments disclosing the collection of location information than health information.  Similarly, the privacy policy segments are heavily skewed towards assertions (i.e. a \textit{positive label}) of data collection than denials (i.e. a \textit{negative label}).

To better characterize the challenge, we tabulate both the number of labeled segments in each category and the percentage of denials for the two data sets and CPPS in Table~\ref{dataset_distribution}. We can see that data for certain categories is extremely sparse: OPP-115 contains thousands of labels on contact data but only $44$ labels on health data. Consequently, Polisis~\cite{harkous2018}, which is trained on OPP-115, struggles with a low accuracy in minority classes such as health. We also see that the data sets exhibit class imbalance across categories, with only 4.19\% of the samples in the OPP-115 having a denial label.  APP-350 has a slightly better class balance, but this is due to the addition of synthetic denials, which they accomplished by manually changing a positive sample into a denial one.  For instance, a positive sample: \textit{``Our App collects your location data''} might be converted to \textit{``Our App does not collect your location data''}.  We note that while this approach mitigates the class imbalance to some extent, it cannot be used to improve data category imbalance.  Furthermore, there is no assurance that such synthetic labels are representative of real denials in the wild. In comparison, after active learning, the CPPS training set generated by \ourtool{} has almost perfect class balance.  As we can see, \ourtool{} is able to find and label many more samples in the 6  rarer data categories.

\begin{table}[]
    \footnotesize
	\centering
	\begin{tabular}{l|l|l|l}
            & OPP-115       & APP-350       & CPPS            \\ \hline
Contact     & 8.46\% | 1217 & 22.5\% | 1744 & 48.07\% | 2852  \\
Location    & 2.45\% | 449  & 22.1\% | 971  & 51.29\% | 2318  \\
Device      & 0.57\% | 700  & 14.9\% | 2062 & 50.57\% | 2828  \\
Demogra.    & 1.35\% | 408  & 17.7\% | 481  & 48.44\% | 3552  \\
Financial   & 5.06\% | 435  & -             & 51.73\% | 2339  \\
Health      & 2.37\% | 211  & -             & 48.60\% | 1790  \\
Survey      & 4.76\% | 84   & -             & 45.13\% | 421               \\
Personal id & 11.3\% | 106  & -             & 38.76\% | 387                \\
Soc. media  & 1.35\% | 74   & -             & 47.15\% | 369                \\ \hline
Average     & 4.19\% | 3684 & 19.3\% | 5258 & 47.75\% | 16856
\end{tabular}\\[-1ex]
	\caption[Data imbalance]{The number of labels across data categories and percentage of denials in previous datasets compared to CPPS\footnotemark}
	\label{dataset_distribution}
\end{table}

\footnotetext{Each privacy policy in OPP-115 was annotated by multiple annotators, thus the number of labels are not equivalent to the number of segments. 0.75 is the middle value of overlap threshold provided by the authors to try to consolidate multiple labels on similar text segments}

\subsubsection{Labeling Cost} \label{challenge:crowdsourcing} 

Previous data sets, such as the OPP-115~\cite{wilson2016} and the APP-350~\cite{zimmeck2019} data sets, are generated with trained human annotators, usually composed of law students.  Even with their prior legal knowledge, it still takes considerable time to read through a privacy policy and label the specific text segments that indicate, for example, whether personal information is collected and if so, what type of information is collected. For instance, OPP-115, which contains 115 privacy policies, was labeled by $10$ law students, spending an average of $72$ minutes on each privacy policy. The speed, availability and cost of trained annotators limits the size of the data set.  Recognizing this challenge, we propose a crowdsourcing solution reduces the cost of labeling by $9\times$. 

One obvious question that arises with the use of crowdsourced annotators for privacy policies is whether such untrained annotators can sufficiently understand privacy policy language to produce accurate labels~\cite{wilson_crowdsourcing_2016}.  We note however, that since privacy policies are the primary method of obtaining consent from individuals for the collection of information, to obtain meaningful consent, they must be understandable by a representative target individual from whom the information collection would likely happen.  For example, Canada's privacy legislation, PIPEDA, states that \textit{``the consent of an individual is only valid if it is reasonable to expect that an individual to whom the organization’s activities are directed would 
understand''} and the GDPR defines consent as \textit{``any freely given, specific, informed and unambiguous indication of the data subject’s wishes''}---in both cases, the user must ``understand'' or be ``informed.''  While it is true that many privacy policy documents fall short of this idealized level of transparency, we argue that the interpretation that an average person has for a privacy policy intended for them (were they to actually read the policy) is still relevant, and is what \ourtool{} tries to predict.  

Nevertheless, one cannot simply use crowdsourced annotators and expect to achieve accurate classification, especially in combination with active learning.  First, crowdsourced annotators acquired through platforms such as Amazon's mTurk service, generally do best on short, simple tasks, such as single-sentence translation~\cite{ambati_active_2010} and sentiment analysis~\cite{chakraborty_asking_2020}.  Further, reducing the complexity and length of tasks is beneficial for crowdsourced annotators~\cite{kittur_crowdforge:_2011, chilton_cascade:_2013}.  Second, active learning methods generally assume reliable oracles that always return correct labels, which is not the case for crowdsourced annotators.  Previous privacy policy studies presented several crowdsourced annotators with the entire privacy policy, asked them each to assign a label, and then attempt to increase label reliability by only accepting the label if the inter-annotator agreement is above some threshold~\cite{wilson_crowdsourcing_2016}. A drawback to this is that different annotators may select different text segments to label, and the text segments themselves may not align perfectly, decreasing inter-annotator agreement.

\ourtool{} addresses these challenges by extracting relevant text segments from a privacy policy and presenting these to the crowdsourced annotators so that all of them label the same segments. This both decreases the time they must spend on the task and increases inter-annotator agreement rate, which both decrease the cost of acquiring labels.  \ourtool{} re-requests labels for text segments that did not achieve an inter-annotator agreement above an acceptable threshold---a mechanism called \textit{relabeling}.  However, some segments may suffer from chronically poor agreement---that is they do not achieve sufficient agreement even after several attempts.  There can be several reasons for the chronically poor agreement for a segment.  First, \ourtool{} may mistakenly extract a segment that is not relevant to the labeling task.  Our crowdsourced task also has workers annotate whether a segment is relevant or not, which is then used as a label to retrain the model used to classify if a text segment contains relevant information about a data category.  Second, the text segment may be relevant but inherently ambiguous---that is the text is not sufficiently clear about whether collection is occurring or not~\cite{reidenberg_ambiguity_2016}. For example, in 
\begin{quoting}
\textit{``if you log-in using a third party social media account, we may not collect any personal or account information from that social media provider''}    
\end{quoting}
the ``may'' makes it ambiguous whether information is collected or not, and users have no control over the usage of private data.  In addition, using the ambiguous term ``personal information''  without further details makes it impossible to infer which data category the segment refers to and undermines the purpose and
value of privacy policies~\cite{reidenberg_ambiguity_2016}. To prevent wasting more crowdsourcer resources on such text segments, \ourtool{} stops the relabeling attempt after a preset number of iterations, and labels such segments as \textit{ambiguous} and does not use them for training.  While classifying all text segments that do not achieve an acceptable level of inter-annotator agreement as ambiguous necessarily over-approximates the true set of ambiguous text segments, we show that \ourtool{} is still able to achieve high classification accuracy despite the loss of potentially usable text segments due to this over-approximation.

Our current implementation of \ourtool{} uses crowdsourced annotators from Amazon's mTurk service to provide labels used for training.  While mTurk captures a large portion of the mobile app user population, who are the target individuals of mobile app privacy policies in our data set, we note that mTurk cannot capture every subgroup of mobile app user.  For example, children are not permitted to use the mTurk service, so our current \ourtool{} models cannot predict how children might interpret privacy policies.  Nonetheless, this is a limitation of mTurk, not \ourtool{}---we believe that our current implementation of \ourtool{} can be extended in a straightforward manner to predict for any subgroup of individuals so long as annotators from that subgroup can be recruited to provide labels from which \ourtool{} can learn.

\subsection{Segmentation} \label{sec:challenge_seg}

As mentioned earlier, extracting text segments is crucial to \ourtool{}'s ability to provide accurate, broad classification of privacy policies at a lower annotation cost. While previous work, such as Polisis, used existing off-the-shelf segmentation tools~\cite{GraphSeg}, \ourtool{} requires text segments for annotation, not just classification. For accurate annotation by crowdsourced workers, \ourtool{} must extract text segments that not only semantically coherent to a topic, but relevant to the specific data category for which an annotation is being sought. As a result, we build custom data category-specific \textit{Segmenters} in \ourtool{}.

A \ourtool{} Segmenter first identifies sentences that are relevant to a data category, and then uses standard topic similarity metrics to \textit{contextualize} the sentence into a coherent text segment. Finally, in contrast to previous work, we also evaluate the safety of using text segment level classification to assign global labels to a privacy policy (Section~\ref{sec:full}). Specifically, we use our Segmenters to examine how often privacy policies may have multiple contradictory text segments, and whether these represent real contradictions where segment-level labels may not be consistent with a privacy policy-level label.

\section{Design Overview}

\begin{figure}[]%
	\centering
	\subfloat[\centering Example with multiple labels on data action and action mode]{{\includegraphics[width=7.6cm]{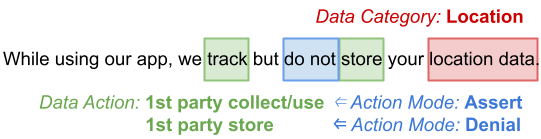} }}\\[-0.5ex]%
	\subfloat[\centering Example with multiple labels on data category]{{\includegraphics[width=7cm]{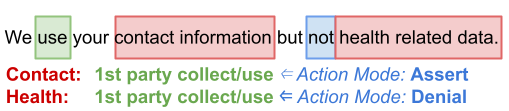} }}\\[-1ex]
	\caption{Privacy Policy segments with multiple labels}%
	\label{fig:multi-labels}%
\end{figure}

The goal of \ourtool{} is to train a hierarchical multi-label classification model that extracts the \textit{action mode} associated with a \textit{data action} for a set of \textit{data categories} from privacy policy text.  \ourtool{} accomplishes by first extracting relevant text segments from a privacy policy and then labeling those segments as shown in Figure~\ref{fig:category_example}.  %
At the highest level, \ourtool{} labels text segments into 9 data categories: \textit{contact, device information, location, financial, health, demographic, survey, personal identifier} and \textit{social media data}.  The next level identifies 3 data actions: \textit{1st party data collection/use, 3rd party sharing} and \textit{data storage}; and then 5 possible action modes are labeled for each data action: \textit{assert} (action is performed), \textit{choice} (action is performed but the user may opt-out), \textit{denial} (denial of the action being performed), \textit{not mentioned} (irrelevant to data practice), and \textit{ambiguous} (identify policy texts that are poorly described, which aligns with interests of recent legal studies~\cite{reidenberg_ambiguity_2016, lie:utlj2021:automated_accountability}). Note that a single text segment may contain information about multiple data categories and multiple data actions, as shown in Figure~\ref{fig:multi-labels}, making this a fine-grained multi-label classification task.

\begin{figure*}[ht!]
	\centering
	\includegraphics[scale=0.32]{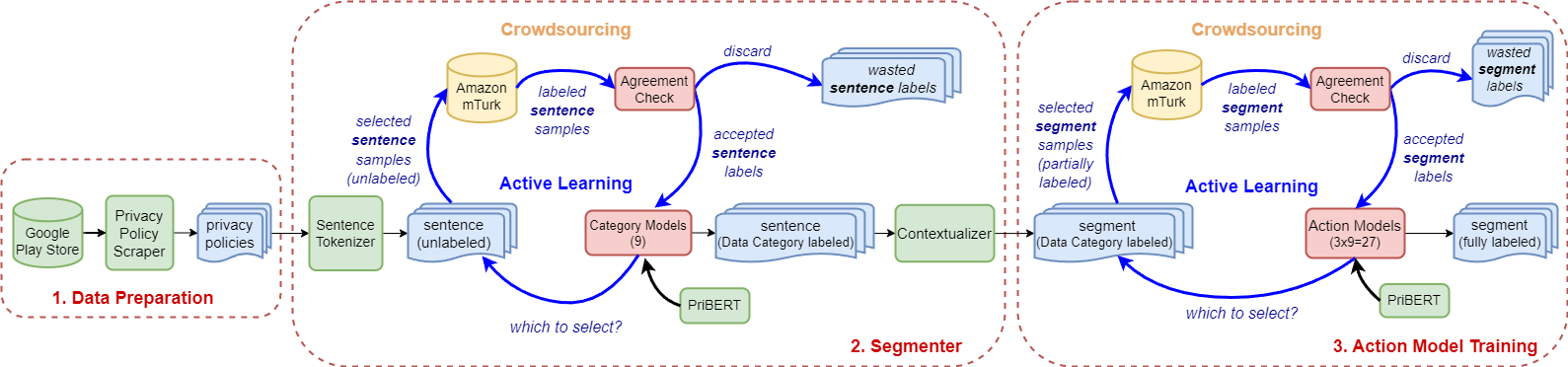}\\[-1ex]
	\caption{A simplified overview of the active learning system}
	\label{fig:overview}
\end{figure*}

\subsection{The \ourtool{} Pipeline} \label{sec:design:pipeline}

\ourtool{} is implemented as a pipeline that processes the input in 3 major stages as illustrated in  Figure~\ref{fig:overview}.  First, a \textit{Data Preparation} stage ingests the raw privacy policies and performs basic pre-processing to prepare them for labeling by the crowdsourced annotators (5 per segment in our study) and subsequent training.  
Second, a \textit{Segmenter} stage extracts text segments relevant to each data category from the processed privacy policies and simultaneously trains and updates the Segmenter models with active learning.
The final \textit{Action Model Training} also uses active learning to train models that produce multi-class action mode labels for each data action/data category combination.

\ourtool{}'s pipeline effectively decomposes the learning task into two separate models.  The Segmenter includes a \textit{Category Model} that selects text segments that are relevant to one of the data categories.  After this, \textit{Action Models} at the next stage only have to predict action mode labels for each data action. This decomposition has two benefits.  First, the decomposition of a single task into two simpler sub-tasks fits the requirements of crowdsourcing, which requires short, simple tasks.  Second, active learning requires an even distribution of bootstrap data, ideally in all combinations, which becomes intractable as the label dimension grows~\cite{ref3}.  Decomposing the labeling task reduces its dimensionality.  We now describe each of the pipeline stages in more detail.

\subsection{Data Preparation} \label{subsec:data_prep}

The Privacy Policy Scraper downloads metadata of $375K$ Android applications from the Play Store and filters down to a final dataset of 51,781 usable privacy policies in the plain text format. The processing details are included in Appendix~\ref{appendix_ppscraper}.

\label{sec:overview:pribert}
\ourtool{} then prepares a privacy-specific contextualized embedding named PriBERT.
Specifically, we perform additional pretraining on the \emph{bert-base-uncased} model using the 52K privacy policies collected by the Privacy Policy Scraper, with a maximum sequence length of 128 and a batch size of 8. Each training checkpoint captures the BERT activations from the last 4 hidden layers of the previous transformer checkpoint. We also apply a truncation rate of $2\%$ to provide robustness to non-sentential inputs. %

\subsection{Segmenter} \label{subsec:seg}
We create data category-specific specific Segmenters by training a model that identifies relevant sentences, and then using standard NLP tools to identify the appropriate segment boundaries.  We begin by using the NLTK~\cite{nltk_tool} Sentence Tokenizer to split entire privacy policies individual sentences.  Bullet lists are converted into a sentence heuristically by concatenating each list item with the text just before the list items.

The individual sentences are passed to a Category Model, which classifies if the sentence is relevant to a data category. There are 9 separate Category Models in \ourtool{}, each responsible for one data category \ourtool{} currently supports and a sentence can be classified as being relevant to more than one data category. Finally, relevant sentences will be \textit{contextualized} by adding additional sentences into a \textit{text segment}, which provides the context necessary for the annotators to understand and apply the correct label. For example, consider the text segment:

\begin{quoting}
	\textit{Personal information is data that can be used to uniquely identify or contact a single person. When you visit, download or upgrade our app or our products, we do not use this information explicitly. However, we may collect personal information to improve our services and deliver a better experience.}
\end{quoting}

All sentences in the segment contribute to its meaning: without the first sentence, it is not clear that personal information refers to contact information, but if the segment were to only contain the first two sentences, annotators may be misled to provide a denial label to the segment.  At the same time, including extraneous sentences in the segment increases the time and effort for annotators, which may increase labeling costs. Sentences are contextualized by the \textit{Contextualizer} component.  We describe the Category Model and Contexturalizer that make up the Segmenter in more detail below.

\subsubsection{Category Models}

For each data category, \ourtool{} trains a binary Category Model to identify relevant sentences: it projects each sentence into a vector representation using the PriBERT embedding mentioned in Section~\ref{sec:overview:pribert}. These vectors are then fed into a set of binary classification models fine-tuned on top of BERT. We set the batch size to 32 and the number of epochs to 48. We use leaky ReLUs~\cite{Maas_Hannun_Ng} and a dropout value ~\cite{Srivastava_Hinton_dropout_2014} of $0.1$ to prevent over-fitting while sustaining the weight updates to avoid vanishing gradient problems in the propagation process. We apply a softmax activation function to the model output and use cross entropy loss and ADAM optimizer~\cite{kingma_adam_2017} to update model weights. 

Each Category Model is initialized using a bootstrap training set of 200 sentences to be labeled.
The bootstrap set is prepared in two steps: first, \ourtool{} randomly selects a number of privacy policies from the unlabeled pool, and crowdsourced workers are asked if the target data category is mentioned in these given documents. While this requires the annotators to read through the full documents, they do not need to give detailed annotations. The privacy policies that are annotated as mentioning a target data category will be tokenized into sentences, and $200$ sentences are randomly selected as the bootstrap set for each of the 9 Category Models. After this, active learning is applied to select more sentences for labeling. It is possible that the bootstrap set does not cover all classes and the model will fail to converge, but we did not encounter this situation in all our experiments. If such a situation arises, it would be necessary to expand the size of the bootstrap set until all classes are represented.

\subsubsection{Contextualizer}

The Contextualizer starts with a relevant sentence selected by the Category Model and iteratively expands it into a text segment by examining the sentences immediately before and after to see if either should be added to the current segment. To do this, the Contextualizer computes the similarity of the sentences to the current text segment using the Word Mover's Distance (WMD)~\cite{Kusner_Sun_Kolkin_Weinberger} and compares it against a threshold calculated from the WMD mean and standard deviation of sentences within the current privacy policy~\cite{Dias_Alves_Lopes}.  This process continues until both the previous and subsequent sentences are not similar enough to be added to the current text segment.

\subsection{Action Models} \label{subsec:cls}

\begin{figure}
	\centering
	\includegraphics[scale=0.38]{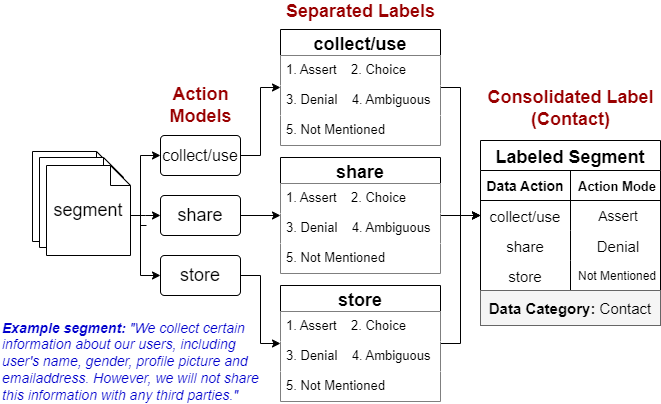}\\[-0.5ex]
	\caption{The action classification process to produce a label on the given \textit{contact} sample segment}
	\label{fig:classifier}
\end{figure}

Action Models are the final stage of the \ourtool{} pipeline. Similar to Category Models, Action Models are designed with three fine-tuned BERT-based classifiers, each responsible for labeling a data action for each data category (9 data categories with 3 data actions for a total of 27 Action Models). Unlike the binary Category Models, Action Models are multi-classed, as shown in Figure~\ref{fig:classifier}.  Together with the relevancy label obtained from the Category Model, the input segment is assigned labels on action modes for each data action.   %
These labels result in a complete segment-level label. We bootstrap Action Models by sending segments output from Category Models to mTurkers to label until there are at least 20 labels for each of the 5 action mode options for the 3 data actions, resulting in at least 300 labels per data category.

\section{Active Learning and Crowdsourcing} \label{sec:AL_and_CS}

\subsection{Crowdsourcing Training Data} \label{section:crowdsourcing}

It is well-known that crowdsourced workers may have varying degrees of skill and may not pay full attention when performing tasks~\cite{costante_sun_petkovi_hartog_2012}, which is especially dangerous for \ourtool{} which relies on active learning to obtain training labels.  To mitigate these challenges, \ourtool{} follows best practices of qualifying mTurk workers as annotators, including specific requirements and qualification tests specific to data category, and designing a simple question survey to perform the labeling task. Details of this are provided in Appendix~\ref{appendix_qual}.  
\textit{The use of human annotators was approved by our institutional review board (IRB).}

After obtaining labels from crowdsourcers, \ourtool{} carefully examines them to ensure label reliability. Only labels with an inter-annotator agreement greater or equal to the \textit{Acceptance Threshold} ($80\%$) will be accepted as training samples. All accepted labels are used in the next training iteration of the appropriate Category Model or Action Model, while unaccepted labels are discarded, referring as the \textit{Wasted Requests} as they still cost money but confer no benefit to the training of \ourtool{}'s models.  \ourtool{}'s Segmenter increases the agreement rate thus decreasing waste, directly leading to reduced labeling costs (Section~\ref{section:eval_seg}).

\label{section:label_correctness}

\subsection{Active Querying Strategy}

As mentioned in Section~\ref{sec:design:pipeline}, \ourtool{} uses two types of classifiers, the Category Models and the Action Models, both of which are trained with active learning.  During active learning, the Category Model selects sentences from the entire unlabeled pool of privacy policy sentences, while the Action Model selects text segments from the pool currently labeled by the Category Model as relevant to the target data category.  As a result, the sentence selected by the Category Model may not be in the text segment selected by the Action Model.  To consolidate the selections and simultaneously train both models, \ourtool{} initially prioritizes the Category Model, otherwise the Action Model may select too many irrelevant text segments for labeling.  Thus, we initially allow the Category Model to pick sentences for labeling until it reaches an F1 threshold of 85\%, and then give the Action Model priority in selecting segments for labeling requests and continue training on both models.

To select samples to label, \ourtool{} uses pool-based sampling as this method is best suited to settings such as ours that have a large unlabeled pool~\cite{settles_2012}. Our implementation builds on top of the \textit{modAL} active learning framework, which is built upon the study of Cardoso et al.~\cite{cardoso_ranked_2017}. Further details are presented in Appendix~\ref{appendix_querying}.  We use Uncertainty-based Sampling~\cite{Lewis94heterogeneousuncertainty}, which selects the samples with the greatest uncertainty for labeling, and modify it for our situation, which has a very large pool of unlabeled samples.  Specifically, we compute not only the uncertainty for the samples in our pool, but also record the history of uncertainty scores.  Samples that achieve a low-level of uncertainty over a period of time are likely never to be selected since uncertainty scores generally decrease as the model learns more about the input space.  As a result, we remove these from the pool to save having to recompute the uncertainty for these on each active learning iteration, as the cost of computing uncertainty for every unlabeled sample is significant due to their large number. 

Note that the active learning process is designed to be online, but as a prototype, we still perform manual checks throughout the process to minimize wasted mTurker funds in case of program bugs. For most iterations, \ourtool{} only updates the model with new labels from mTurkers, which takes <1 min. To prevent catastrophic forgetting, we retrain all labels after 10 iterations. The main bottleneck is waiting for mTurkers to accept HITs and provide annotations.

\subsection{Relabeling Strategies} \label{subsec:design_relabel}

Recall that text segments whose inter-annotator agreement fails to cross the agreement threshold are not accepted for training and waste labeling resources.  \ourtool{} currently discards these segments, a  policy, we call \textit{Label and Discard}. Label and Discard protects the trained model from training on incorrectly labeled samples, which can lead to lower accuracy, as shown in our Evaluation in the \textit{Acceptance Threshold} section in Appendix~\ref{evaluation_sensitivity}.  \ourtool{} aims to add 30 new labels each active learning iteration before selecting another set of samples to label.  Since, we know all unaccepted labels will be wasted, we can estimate the number of label requests we need to make to attain 30 accepted labels.  We have measured the acceptance rate for our survey to be approximately 73\%, which requires \ourtool{} to request $42$ surveys each iteration.  

Another option, rather than wasting the unaccepted labels, is to re-request them to see if the agreement rate for a larger number of annotators can be acceptable.  We call this policy \textit{Incremental Relabeling}, which we implemented based on that of Zhao et al.~\cite{zhao_sukthankar_sukthankar_2011}. Incremental Relabeling trades off the possibility of recovering some of the wasted labeling resources for the risk of wasting even more resources.  We put a cap of 3 on the number of times any text segment can be requested for labeling for a total of up to 15 annotations and keep the agreement threshold at 80\% (i.e 4/5, 8/10 or 12/15).  We evaluate these two policies in the \textit{Relabeling} section in Appendix~\ref{evaluation_sensitivity}.

\section{Evaluation and Measurements} \label{section:evaluation_complete}

We begin by evaluating the accuracy of \ourtool{}'s Category Models (9) and Action Models (27) against a test set of segments whose labels have been provided by law students. We then evaluate the accuracy of the Segmenter against segments that were labeled by those trained annotators, as well as the Segmenter's contribution to reducing annotation cost. We also evaluate \ourtool{}'s use of active learning to reduce annotation costs and mitigate data imbalance.  Finally, we compare the cost of \ourtool{}'s use of crowdsourced annotators versus the cost of using trained annotators. Measurements of \ourtool{}'s sensitivity to the acceptance threshold, relabeling strategy and size of the unlabeled pool are given in Appendix~\ref{evaluation_sensitivity}.

\subsection{Accuracy of Category and Action Models}\label{sec:eval:classification} 

We measure the absolute accuracy of \ourtool{}'s predictions against the law student-labeled test set. We then show the benefits of training on the CPPS dataset by comparing \ourtool{}’s Category and Action Models against a baseline classifier trained on OPP-115 and APP-350. 

\subsubsection{Absolute Accuracy}  \label{sec:full}

As APP-350 and OPP-115 do not have labels for ``choice'' or ``ambiguous'' we set aside 30 labeled segments per data action in each data category for a total of 810 test segments. We then train \ourtool{} without the test segments and tabulate \ourtool{}'s accuracy in Table~\ref{table:end-to-end}. To save space, we show the accuracy of each action mode, averaged across all three data actions in each data category, as well as an overall accuracy for each data category.

The evaluation shows that \ourtool{} can achieve good accuracy across both common and rare data categories, as well as good accuracy in all the fine-grained action model labels. Specifically, it achieves an average f1 of 0.86 and an overall accuracy of 0.90, with the highest f1 in \textit{Assert} (0.92) and \textit{Not Mentioned} (0.91), and the lowest in \textit{Ambiguous} (0.78). We suspect this is because \textit{Assert} statements are usually described clearly as per the legal requirements, while \textit{Not Mentioned} labels are very different from labels of the remaining 4 action modes which all have some contextualized relationship with the target data category. Therefore intuitively it is easier for \ourtool{} to correctly predict on \textit{Not Mentioned} labels. On the other hand, its accuracy in predicting \textit{Ambiguous} labels is slightly lower, as there are many factors leading to ambiguity in privacy policies, e.g. long sentences, terms being too general, confusing wordings, incomplete information, and so on. However, we believe that an F1 score of 0.78 is still very promising, given the fact of the potential noise within the ambiguous labels.

\begin{table}[]
\footnotesize
\centering
\begin{tabular}{l|llllll}
\textbf{data category} & \textbf{denial} & \textbf{assert} & \textbf{choice} & \textbf{amb.} & \textbf{NM}  & \textbf{avg. f1} \\ 
\hline
contact                                          & 0.80                                    & 0.93                                      & 0.83                                      & 0.78                                      & 0.94                                    & 0.86                         \\
location                                         & 0.79                                    & 0.85                                      & 0.77                                      & 0.67                                      & 0.94                                    & 0.80                         \\
device                                           & 0.89                                    & 0.92                                      & 0.83                                      & 0.82                                      & 0.96                                    & 0.88                         \\
demographic                                      & 0.81                                    & 0.88                                      & 0.86                                      & 0.80                                      & 0.86                                    & 0.84                         \\
financial                                        & 0.88                                    & 0.96                                      & 0.80                                      & 0.82                                      & 0.86                                    & 0.86                         \\
health                                           & 0.94                                    & 0.95                                      & 0.82                                      & 0.89                                      & 0.96                                    & 0.91                         \\
survey                                           & 0.82                                    & 0.90                                      & 0.80                                      & 0.78                                      & 0.91                                    & 0.84                         \\
personal id                                      & 1.00                                    & 1.00                                      & 0.97                                      & 0.82                                      & 0.97                                    & 0.95                         \\
social media                                     & 0.81                                    & 0.90                                      & 0.93                                      & 0.67                                      & 0.81                                    & 0.82                         \\ \hline
average                                          & 0.86                                    & 0.92                                      & 0.85                                      & 0.78                                      & 0.91                                    & 0.86       
\end{tabular}\\[-0.5ex]
\caption{F1 score on the CPPS test set. We show the f1 for each of the denials, assertions, choice, ambiguous (amb.) and not-mentioned (NM) labels, as well as the average f1 value.}
\label{table:end-to-end}
\end{table}

\subsubsection{Benefits of CPPS}\label{sec:eval:classifier_selector}

As explained in Section~\ref{challenge:breadth_fine_classification}, previous tools classify at a coarser granularity than \ourtool{}, so we cannot directly compare the accuracy of labels produced by \ourtool{} against theirs. As previously shown in Table~\ref{tab:tools_comparison}, previous tools label privacy policy text segments into two classes, while \ourtool{} provides 5. However, all previous tools were trained exclusively on the APP-350 and OPP-115 data sets, while \ourtool{} is trained on the larger and balanced CPPS dataset that was produced by \ourtool{}'s training approach. It is therefore unfair to compare \ourtool{} against existing tools. Instead, we train baseline models with identical architectures as those of \ourtool{}'s on data from APP-350 and OPP-115. We randomly select some segments from the the APP-350 and OPP-115 as the test set and tabulate the F1 result in Table~\ref{tab:baseline_rerun}. \ourtool{} outperforms the baseline models on fine-grained classification tasks as well as the overall average across all data categories.

\begin{table*}
\small
\centering
\begin{tabular}{ll|lllllllll|l}
           &          & contact       & device        & location      & demogr.       & financial     & health        & survey        & pers. id      & soc. media    &  average              \\ 
\hline
Category   & Calpric  & \textbf{0.94} & 0.90          & \textbf{0.89} & \textbf{0.97} & \textbf{0.94} & \textbf{0.89} & \textbf{0.87} & \textbf{0.78} & \textbf{0.82} & \textbf{0.89}  \\
           & baseline & 0.90          & 0.79          & 0.85          & 0.78          & 0.92          & 0.43          & 0.80          & 0.63          & 0.75          & 0.76           \\ 
\hline
Action & Calpric  & \textbf{0.97} & \textbf{0.91} & \textbf{0.92} & \textbf{0.90} & \textbf{0.89} & \textbf{0.98} & \textbf{0.96} & \textbf{0.84} & \textbf{0.89} & \textbf{0.92}  \\
           & baseline & 0.91          & 0.68          & 0.62          & 0.65          & 0.63          & 0.54          & 0.60          & 0.57          & 0.57          & 0.64          
\end{tabular}\\[-1ex]
\caption{F1 comparison between \ourtool{} Category and Action models vs baseline model}
\label{tab:baseline_rerun}
\vspace{-12pt}
\end{table*}

\subsection{Segmenter} \label{section:eval_seg}
In previous tools, text segments used for training were extracted by human annotators while \ourtool{}'s training approach automatically extracts text segments. We thus measure how closely \ourtool{}'s segments extracted by \ourtool{}'s Segmenter cover the same segments extracted by  human annotators. To show that the Segmenter increases inter-annotator agreement, we compare the level of agreement with and without automated text segmentation. Finally, we investigate the frequency and impact of conflicting text segments in privacy policies.

\subsubsection{Segment Coverage}

To examine the ability to capture useful segments from privacy policies, we compare the outputs of \ourtool{}'s Segmenter against the text segments extracted by human annotators in APP-350. We measure the number of labeled sentences in APP-350 in the relevant data categories that are covered by (i.e. appear entirely in) some segment extracted by our Segmenter.  We run our Segmenter on all the original privacy policies provided by APP-350 and find that $97\%$ of the sentences in APP-350 are covered by our Segmenter.  We examine the uncovered sentences and find that the vast majority are either ambiguous or irrelevant sentences.  For example, in APP-350, the following segment is labeled as collecting contact information:
\begin{quoting}
\textit{``Anonymous Data AI Factory may collect and use anonymous data, via Google Analytics. We cannot identify you from this data. Such data is pooled to give us general statistics on our audience and usage of our apps.''}
\end{quoting} 
but is not labeled as collecting contact information by our Segmenter.

\subsubsection{Increasing Inter-annotator Agreement}\label{sec:eval:sl_vs_fdl}

We now evaluate the benefit of having annotators label text segments instead of existing approaches such as full-text or paragraph-level labeling. We randomly select $3$ privacy policies from APP-350 and conduct an experiment where mTurk annotators are asked to either label the full document, paragraphs produced by existing preprocessors, or text segments extracted by \ourtool{}. For full document labeling (FDL), we present the full document to annotators and ask them to label 20 sentences that denote any of the data actions and action modes on any of the relevant data categories. For paragraph labeling (PL), we present paragraphs for annotators to select and label. We generate paragraphs using a modified version of the PolicyLint preprocessing script. We exclude outlier paragraphs falling out of the middle two quartiles (i.e. 25-75\%) of the distribution in terms of length, as they might be excessively long or so short as to contain no useful information. For segment labeling (SL), we have annotators label 20 segments per privacy policy using the \ourtool{}'s survey. To align with our labeling scheme (i.e. includes more classes), which differs from the original APP-350, selected samples are reviewed and re-labeled by law students. We calculate label accuracy (Acc.) by measuring the percentages of raw labels that are correctly labeled by mTurkers. A label is considered accurate if and only if it covers one of the labeled sentences in APP-350 and the labels in data category, data action and action mode all match.  

The results are tabulated in Table~\ref{table:segmentation_tool}.  We found that SL produces the most efficient results: annotators needed roughly 4 times and 2 times as long to produce the same number of labels using FDL and PL, respectively. That is, 21 minutes for SL, 46.5 minutes for PL, and 90 minutes for FDL (per policy). As a result, to maintain the same hourly pay rate, FDL and PL require 4$\times$ and 2$\times$ the payment to obtain the same amount of labels as using SL. We find that SL gives the highest accuracy. Closer inspection of FDL shows that some annotators incorrectly label segments such as \textit{``We do not track your location unless we have your opt-in consent.''} as ``No'' instead of ``Choice'', as they truncated the phrase after ``location'', thus missing the later part that makes it ``Choice''. SL also outperforms PL by 15\%. We found that this was mainly because text critical to understanding a data action may span a paragraph break.

We also observe that the agreement rate (AR) for FDL is significantly lower, as annotators often disagree on text boundaries for their labels.  This result aligns with what previous scholars observed~\cite{wilson2016}, even when labels are created by trained annotators. On the other hand, ARs for PL and SL are similar, with PL being slightly higher (1.3\%).

We compute the raw cost per label by dividing the total payments by the number of unique labeled segments, the effective cost after discarding unaccepted segments, and the time taken to produce each raw label. FDL is both more expensive (4$\times$ for raw labels, and 5.2$\times$ for accepted labels), and requires 4.3$\times$ as much time to produce raw labels than SL.  Qualitatively, we also found that because annotators are free to select the text they wish to label, they tended to select easier segments that could be found by simply searching the text for keywords (i.e. ``demographic'' or ``contact'') instead of carefully reading the complete privacy policy.  As a result, this may exacerbate class imbalance and lead to poorer coverage of the privacy policy text. PL has a better performance than FDL, yet is still 2.2$\times$ more expensive than SL.

\begin{table}
\footnotesize
\centering
\begin{tabular}{|l|l|l|l|l|l|} 
\hline
             & \textbf{Acc.} & \textbf{AR} & \textbf{c\_raw} & \textbf{c\_accept} & \textbf{t\_raw}  \\ 
\hline
\textbf{FDL} & 55\%          & 6.7\%       & \$0.5           & \$4.55             & 4.5 mins         \\ 
\hline
PL           & 63.33\%       & 73\%        & \$0.28          & \$1.88             & 2.32 mins        \\ 
\hline
\textbf{SL}  & 78.33\%       & 71.67\%     & \$0.125         & \$0.87             & 1.05 mins        \\
\hline
\end{tabular}\\[-1ex]
\caption{Comparison between Full-Document Labeling (FDL), Paragraph Labeling (PL), and Segment Labeling (SL) using crowdsourcing: label accuracy(Acc.), agreement rate (AR), the average cost per raw label (c\_raw), average cost per accepted label (c\_accept), and the average time took to produce one raw label (t\_raw).} 
\label{table:segmentation_tool}
\end{table}

\subsubsection{Intra-document conflicts} \label{sec:intra-doc-conflicts}

One potential risk of segment-level labeling is that it becomes unclear how to consolidate contradictory labels if more than one relevant segment for a data category and data action appear in the same privacy policy. We examined 115 privacy policies and find that 12\% have contradictory labels for a data category and data action.  Of these, 8\% have conflicts that take the form of a denial that they collect a certain type of data, followed by a statement that they will share that data if legally required to do so. We also compare contradictions found by \ourtool{} with PolicyLint\cite{policyLint}, on OPP-115. PolicyLint finds 20 of the privacy policies have contradictions, 8 of which are also found by \ourtool{}. Of the 12 not found by \ourtool{}, PolicyLint considers 5 contradictions only because it lacks a ``Choice'' label, 2 are due to PolicyLint and \ourtool{} having different subsumptive relationships (i.e. PolicyLint considers demographic information as a type of personal information while \ourtool{} considers them separate classes), 2 because of differences in the annotation scheme and 3 due to differences in the data categories covered. \ourtool{} also finds 6 contradictions that PolicyLint does not find. We manually verified that they are real contradictions and were missed by PolicyLint as it is unable to differentiate between choices and explicit denials.

\subsection{Active Learning} 

The following evaluations demonstrate how active learning enables \ourtool{} to 1) achieve lower labeling costs by reducing the number of labeled training samples, and 2) improve class balance to achieve higher accuracy. 

\subsubsection{Labeling Savings} 

\begin{table}[]
    \footnotesize
	\centering 
\begin{tabular}{l|lllll}
                      & \textbf{n\_nonAl} & \textbf{n\_Al} & \textbf{Al\_save} & \textbf{m\_start} & \textbf{m\_end} \\ \hline
\textbf{contact}      & 1596              & 942            & 40.98\%           & 4.92\%            & 49.68\%         \\
\textbf{location}     & 840               & 632            & 24.76\%           & 7.04\%            & 50.00\%         \\
\textbf{device}       & 998               & 610            & 38.88\%           & 4.20\%            & 46.07\%         \\
\textbf{health}       & 732               & 529            & 27.73\%           & 5.56\%            & 42.53\%         \\
\textbf{finanical}    & 982               & 793            & 19.25\%           & 6.80\%            & 49.94\%         \\
\textbf{demogra.}     & 1428              & 1149           & 19.54\%           & 7.55\%            & 58.40\%         \\
 \hline
\textbf{average}         & 1096              & 776            & 28.52\%           & 6.03\%            & 49.44\%        
\end{tabular} \\[-1ex]
	\caption{Comparison between active vs. non-active Category Models: $n$ is the number of training labels required to achieve F1=0.85; A higher $AL\_save$ indicates a larger improvement in the learning progress; $m$ is the percentage of samples in a minority class, recorded at the bootstrap ($m\_start$) stage and the completion of training ($m\_end$).}
	\label{table:discriminator_eval}
\end{table}

We evaluate the reduction in the amount of labeled training data that active learning provides to \ourtool{}.  To do this, we set up an experiment that evaluates how many training samples are required to achieve a certain Category Model accuracy across all 6 data categories.  We prepare a set of balanced test and training sets for each data category by using samples from APP-350 (and OPP-115 if APP-350 does not label such categories).  Text segments that are labeled as in a data category are assertions, and segments not in the category are denials. We simulate a non-active training procedure by randomly selecting samples from this pool and compare this to a training procedure that uses active learning to select samples. We record the number of training samples required to achieve an F1 score of 0.85 in the non-active learning and active learning cases in the n\_nonAL and n\_AL columns respectively, and the percentage of samples saved in the active learning case in the AL\_save column in Table~\ref{table:discriminator_eval}. We did not include \textit{survey}, \textit{personal identifier} or \textit{social media} because the non-AL model is not able to find enough training samples to achieve the required accuracy. For the other categories, we observe that the average savings in the number of samples required is $28.52\%$. The savings are higher in categories where more samples are required in the non-active learning case, suggesting that there are many similar samples that the random selection in the non-active learning case selects over and over again.  However, this trend is weaker in financial and demographic categories.  Upon further investigation, we observe that these two categories exhibited longer samples with more complex features. As an example, device-related sentences such as 
\begin{quoting}
\textit{``Our app may access your unique device identifier to tailor ads.''}
\end{quoting}
 are usually simple and concise, whereas a demographic-related sentence like
\begin{quoting}
\textit{``App A provides us with access to certain information about such User as is stored in the User's App A account, namely, the User's public profile (i.e., User's name, profile picture, email address, gender and other public information) and his/her list of friends and/or any other information which is detailed and displayed to the User in the notice which appears during the log in process.''}     
\end{quoting} 
is not. This complexity means that a good number of samples are required to achieve higher accuracy regardless of how the samples are selected, which makes the advantages of active learning in these cases, though still significant, somewhat lower than in the cases with simpler samples.  The contact and device categories exhibit the greatest savings.  We believe this is a result of those categories having a heavy-tailed distribution, the tails of which active learning is much better able to target than random selection.

\subsubsection{Class Balance}

To verify that \ourtool{} mitigates class imbalance, we measured class balance as a percentage of samples in a minority class, recorded at the bootstrap phase and at the end of model training, respectively. Table~\ref{table:discriminator_eval} also shows the starting (m\_start) and ending (m\_end) percentages of minority samples in all data categories.  We see that in all categories, the minority class initially starts of low, but eventually approaches close to making up half of the training set ($50\%$) used to train the model.  We also observe an inverse correlation between the increase in minority percentage and the AL saving percentage. This is intuitively reasonable because boosting from a lower minority percentage to a higher minority percentage naturally requires more training samples.   We performed a similar evaluation on Action Models, and found similar results in the improvement of class balance.  In these cases, the increases in the minor classes were 29.6\% in the health, 19.4\% in the financial, and 24.0\% in the demographic categories.

We also compare \ourtool{} with standard upsampling methods to mitigate to class imbalance, such as duplicating or synthesizing minority samples~\cite{johnson_survey_imbalance_2019}. To do this, we randomly select 800 labeled text segments and sort them by whether they are in a minority or majority class. We then duplicate, synthesize or select minority labels with active learning until the training sets are balanced, train our models and compare accuracy. To duplicate samples, we randomly select samples from the minority set and repeat them in the final training set. To synthesize minority samples we randomly select majority samples and negate them using the method described in~\cite{negation_2015}, implemented with Stanza\footnote{\url{https://stanfordnlp.github.io/stanza/}}. Finally, we use active learning to select enough additional minority samples to balance the training set.  We then 
train a fine-tuned BERT model with a batch size of 32  for 48 epoches and evaluate the resulting model on a test set with a naturally occurring distribution of majority and minority samples. We repeat the experiment 5 times and tabulate the loss, accuracy, F1 values for majority and minority test sets, weighted F1, and balanced F1 for the 3 data set balancing methods, as well as training with no balancing method in Table~\ref{table:baseline}.  As we can see, \ourtool{} with active learning achieves better results except for loss. Using no balancing method (None) achieves the lowest loss, but as expected, suffers from low accuracy in the minority class, as it has simply optimized for the majority class.

\begin{table}[]
\small
\centering
\begin{tabular}{l|llll}
Balance alt. & acc.          & F1 (maj/min)                 & wgt. F1       & bal. F1        \\ 
\hline
None         & 0.79          & 0.88, 0.08                   & 0.69          & 0.44           \\
Duplication  & 0.78          & 0.87, 0.29                   & 0.73          & 0.58           \\
Negation     & 0.79          & 0.88, 0.32                   & 0.75          & 0.62           \\
Calpric      & \textbf{0.80} & \textbf{0.89}, \textbf{0.56} & \textbf{0.81} & \textbf{0.73} 
\end{tabular}\\[-1ex]
\caption{Performance comparison against other alternatives, including the non-AL baseline method (Random), upsampling with Duplication and Negation. Metrics include accuracy, F1 for majority and minority class, weighted F1, and balance F1.}
\label{table:baseline}
\end{table}

\subsection{mTurkers Vs. Law students} \label{section:eval_labelers} 

\begin{table}[]
	\centering
 \footnotesize
\begin{tabular}{|l|l|l|l|l|}
\hline
             & AR   & avg\_agree & Acc. & cost \\ \hline
Law students & 0.83 & 0.91       & 0.99 & \$8.20     \\ \hline
mTurkers     & 0.81 & 0.88       & 0.98 & \$0.92     \\ \hline
\end{tabular}\\[-1ex]
\caption{Comparison between labels created by mTurkers vs. law students: acceptance rate (AR), average agreement rate for all questions among all annotators (avg\_agree), label accuracy (Acc.), and the average cost to produce one accepted label (cost). }
\label{table:crowd_vs_experts}
\end{table}

While trained annotators, such as law students, might cost more than crowdsourced annotators, such as mTurk workers, they might also be more effective at generating usable labels.  To investigate this, we evaluate how law students compare to mTurkers at performing the surveys used in \ourtool{} label requests.  We prepare $600$ surveys from segments that are randomly selected from OPP-115 and evenly distributed for each data category (we use OPP-115 instead of APP-350 because APP-350 does not cover as many data categories). We assign the task to $10$ mTurkers and $10$ law students, which is similar to the annotation setup of OPP-115~\cite{wilson2016}.  We pay the law students $\$31$/hr\footnote{We initially advertised a pay rate of $\$20$/hr but increased it after we were only able to recruit 3 law students over a 4 week period.  A quick search on the Internet confirmed that $\$31$/hr is not unreasonable for law student interns.} and use an acceptance threshold of 8/10.  We tabulate the results in Table~\ref{table:crowd_vs_experts}, which shows that the acceptance rate of the law students is slightly higher but still quite similar to that of the mTurkers (83\% vs 81\%).  Similarly the average inter-annotator agreement rate is slightly higher for the law students but still essentially the same.  As a result, law students and mTurkers produce almost the same number of accepted labels per label request, resulting in a cost per accepted label disparity of roughly 9$\times$ in favor of the mTurkers.  To evaluate accuracy, we treat the labels in the OPP-115 set as the correct labels and compute the accuracy of accepted labels for the two types of annotators.  We find that the both mTurkers and law students have almost perfect labeling accuracy ($98\%$ vs $99\%$).

\section{Android App Privacy Policy Analysis} \label{section:policy_analysis}

With \ourtool{}'s capabilities, we study Android App privacy policies by posing three questions as described below. We limit our study to the top 5 app categories (education, business, entertainment, music \& audio, tools and lifestyle) as tracked by AppBrain\footnote{\url{https://www.appbrain.com/stats/android-market-app-categories}}, and add 3 minority app categories: lifestyle, shopping and health \& fitness. We selected these minority categories as we believe they are more likely to use some of the minority data categories included in \ourtool{} currently, such as social media, financial and health.

\vspace{6pt}

\noindent\textit{\textbf{Q1: Is there a correlation between explicit denials and app popularity?}}
We use the number of downloads and app rating as indicators for popularity and tabulate the average number of explicit denials per application in Tables~\ref{tab:analysis-q1-downloads} and \ref{tab:analysis-q1-rating} (in  Appendix~\ref{appendix_app-analysis}).  When considering rating, we remove apps with fewer than 500 reviews and 50K downloads as some apps with few users can have very high ratings. Regardless of whether we infer popularity by rating or downloads, we see that generally apps that are more popular have more denials. However, this trend does not hold for the most popular apps. Upon closer inspection, we find that apps in the most popular category are generally well-known apps supported by large companies and organizations, and as such support a wide range of functionality, resulting in more handling of private information. For instance, Facebook, a representative example of such an app, contains social media functionality, and has functions associated with personal and demographic information. As such, they cannot have explicit denials of collecting and using this information.

\vspace{6pt}

\noindent\textit{\textbf{Q2: Which data actions have more of the rare explicit denial and choice labels?}}
We tabulate the frequency of assertions, denials and choice labels by data action in Table~\ref{tab:analysis-num-labels}. From this, we can see that overall, assertions are the most frequent,  followed by explicit denials, and lastly choice labels. This is reasonable as the primary function of privacy policies is to disclose how they handle private user data. We also find that sharing has the highest number of denial and controllable statements, suggesting that app developers may feel that users are most concerned about this data action. Sharing is often associated with non-critical functionality, such as advertising, analytics or social media, and thus can be excluded more or made controllable by the user to make a potentially useful app more privacy protective.  Curiously, there are more denials of storing data than assertions of it, suggesting that storage may be often implied as opposed to explicitly stated.

\vspace{6pt}

\noindent\textit{\textbf{Q3: Are rarer data categories uniformly distributed among app categories? How does the distribution of rare data categories compare against previous data sets?}} We tabulate the normalized frequency of data category disclosures for each app category in Table~\ref{tab:analysis-datacate-appcate} in Appendix~\ref{appendix_app-analysis}. In summary, common data categories (contact, device, location) are used more or less uniformly across all app categories. However, the distribution of rare data categories is less even. For example, health data is disproportionately used by health \& fitness apps, and shopping apps are more likely to use financial data. As a result, we conclude that rarer data categories are more specialized and tend to be used only by specialized apps.

Second, we find that data categories that did not receive as much attention as they were perceived to be rare, are actually not so rare. We see that the use of social media data, financial data and demographic data is fairly common. We believe that lack of attention to these other categories may have been partially caused by some of the early privacy policy datasets being drawn from web pages instead of mobile apps (such as OPP-115), which resulted in under-representation and poor accuracy for those data categories~\cite{harkous2018}.

\section{Limitations and Threats to Validity}
Because \ourtool{} relies on crowdsourced labels, it cannot account for policy jargon having inconsistent meanings across annotators. For example, sharing has specific meanings in CCPA\footnote{\url{https://oag.ca.gov/privacy/ccpa}}, which may differ across jurisdictions and that crowd-sourced annotators may not be aware of. We note that our evaluations against OPP-115 and APP-115 do not take into account changes in laws that have taken place between the time those datasets were collected and the time we developed \ourtool{}. As \ourtool{} is trained on segment texts, it does not take into consideration definitions appearing in policy headers. Similarly, it cannot take into account text that is not in the privacy policy, such as consent requests in the application itself. While \ourtool{} currently only works on Android privacy policies, we believe the same approach can be used on policies from other sources, such as websites and IoT devices. Finally, Calpric does not have a hierarchy of data categories like PolicyLint/PoliCheck\cite{policyLint,PoliCheck} does, causing some contradictory labels (i.e. collection of phone number and denial of collecting emails, which are both contact information). We believe \ourtool{} can support hierarchical categories and leave this for future work.

\section{Related Work}

\textbf{Privacy Policy Analysis: }In contrast to previous studies, we present a system that automatically collects diverse training samples before labeling and support precise information extraction and summarization on a broad range of categories. %
Most existing automated systems analyze privacy policies in a coarse granularity, focusing on one-level categorical taxonomy. Watanabe et al.~\cite{watanabe_2018} used keyword-based search model to identify non-compliance between mobile apps and their privacy policies. Costante et al.~\cite{costante_sun_petkovi_hartog_2012} presented a solution to automatically assess the completeness of a policy using more complex algorithms such as Linear Support Vector Machines (LSVM). Wilson et al.~\cite{wilson2016} created OPP-115, a data set of 115 website privacy policies labeled by legal experts. Frederick et al.~\cite{Liu_Wilson_Story_Zimmeck_Sadeh} presented a performance comparison among Logistic Regression (LR), Support Vector Machine (SVM), and Convolutional Neural Network (CNN) models trained on OPP-115, focusing on the categorical levels only. Zimmeck et al.~\cite{zimmeck_automated_2017} developed a document-level compliance checking tool that focuses on 9 privacy requirements, covering 3 data categories, 2 data actions and 2 notices. Recent works began to explore sentence or segment-level classification in privacy policies. Harkous et al.~\cite{harkous2018} developed a multi-label classifier using a CNN to label policies using the OPP-115 taxonomy, which covers both the top-level categories and the second-level attributes. However, 
they suffer from low accuracy for minority classes such as health due to lack of training samples. On the other hand, Zimmeck et al.~\cite{zimmeck2019} created a labeled mobile app-specific privacy policy corpus, APP-350, and trained SVM classifiers to analyze privacy requirements. To compensate the rarity of denial annotation labels, they created synthetic data by manually changing assertions into denials. In contrast, we overcome class imbalance by mining a large unlabeled dataset for potential denial samples and getting them labeled with active learning.

Wilson et al.~\cite{wilson_crowdsourcing_2016} investigated and confirmed the viability of extracting labels from privacy policies through crowdsourcing. While they focus solely on crowdsourcing methodologies, we extend our work to the classification of privacy policy. Despite crowdsourcing being a cheaper option, on average, they reported a cost of \$60 to label each privacy policy. Our system is able to further reduce the average cost of \$13.5. Zimmeck et al.~\cite{Zimmeck2014} also explored crowdsourcing, but did not train their ML models on these labels. They did, however, bring up an important issue of privacy policies collected via crowdsourcing: the low inter-annotator agreement among labels. In both prior works, workers are given an entire privacy policy. To create a label, they first need to read through the document, select and highlight a segment of texts among all texts, and then do detailed annotation based on the specific taxonomy. Because workers may select texts differently based on their own interpretations, it is difficult to measure the inter-annotator agreement and filter out labels with low confidence. To overcome this issue and further reduce the labeling cost, we simplify the crowdsourcing tasks by designing a novel segmentation algorithm and survey-based tasks. PolicyLint~\cite{policyLint} and PoliCheck~\cite{PoliCheck} use an ontology to organize information types as a hierarchy and identify conflicts caused by coarse-grained labels, similar to what we discussed in Section~\ref{sec:intra-doc-conflicts}. However, instead of supervised deep learning, they employ statistical name-entity recognition.

\noindent\textbf{Active Learning:} 
In general, there are three active learning scenarios: membership querying synthesis, stream-based, and pool-based selective sampling. Pool-based selective sampling has been the most well-studied scenario, especially for text classification~\cite{Tong_Koller} and information extraction~\cite{Settles_Craven_Friedland}. The major advantage of this method is that it evaluates and ranks the entire set of unlabeled training points before selecting the next one to label~\cite{settles_2012}. We select pool-based sampling because it is applicable to \ourtool{}'s scenario as we have the entire unlabeled training set up front and it is the most effective and widely used sampling mode for text classification.

To our knowledge, \ourtool{} is the first deep active learning classifier for privacy policies. There are related studies using similar learning approaches in areas such as image analysis and classification~\cite{bio_AL,gal_deep_2017}. Zhang et al.~\cite{zhang2016active} implemented active learning strategies on top of CNNs for sentiment analysis, whereas Shen et al.~\cite{shen_2017_deep} investigated uncertainty-based active learning heuristic for sequence tagging on a newly proposed CNN-CNN-LSTM architecture. We are also the first paper that explores active learning on multi-label BERT models~\cite{devlin2018bert}. Although there are recent papers studying active learning on the, they focus solely on single-label problems~\cite{grieshaber_fine-tuning_2020, ein-dor_active_2020}, where each data is restricted to have one label.

\noindent\textbf{Unreliable Oracles:} Most existing solutions on unreliable oracles in active learning require a confidence score for the oracle and also assume there exist at least some reliable oracles to use (though perhaps at a higher cost)~\cite{ref1}. In other words, the assumption is still impractical especially in the setup where annotation tasks are complex and time-consuming. A more recent study realizes the fact that the noise in human oracles may be non-uniformly distributed, yet the solution still relies on the confidence level~\cite{ref2}. In privacy policy classification, when using crowdsourced labels, it is difficult to derive an accurate estimation of the confidence score. There are also studies on crowdsourcer reliability and crowdsourced label quality~\cite{rouse_reliability_2015, ref2, lovett_data_2018}. However, they either do not involve machine learning or do not apply to an active querying setup. 

\section{Conclusion}

In this work, we simultaneously address both problems of cost and data imbalance through a combination of automatic text selection and segmentation, crowdsourcing and active learning.  In analyzing the properties of our solution, \ourtool{}, we find all three components are essential to its success.  Automatic text selection and segmentation are crucial to be able to reliably use crowdsourced annotators, as shown by our comparison of automatically segmented labeling tasks with other alternatives used by previous work (Section~\ref{sec:eval:sl_vs_fdl}).  Enabling the use of crowdsourced annotators gives a 9$\times$ saving in cost over trained annotators such as law students (Section~\ref{section:eval_labelers}).  Finally, having cheaper, reliable on-demand annotators enables the use of active learning, which further contributes to decreasing the cost by selecting a more optimal set of segments for labeling to achieve greater accuracy and improved balance across data categories and action modes.  
\section*{Acknowledgments}
We thank the shepherd and anonymous reviewers for their feedback. This research was supported by a Tier 1 Canada Research Chair in Secure and Reliable Systems, Telus Corp, and NSERC Grants RGPIN-2018-05931 and CRDPJ 535902-18. Wenjun Qiu is partially supported by an SRI fellowship.

\bibliographystyle{plain}
\bibliography{ref}

\appendix

\section{Privacy Policy Scraper} \label{appendix_ppscraper}

\ourtool{} requires a small set of labeled privacy policy samples to bootstrap the active learning process and a large set of unlabeled privacy policy samples as the potential training set. %
The Scraper downloads metadata of $375K$ Android applications from the Play Store, eliminates broken links by checking the HTTP return status, and filters out non-legal documents with keyword checking heuristics~\cite{harkous2018, HosseiniDegelingUtzHupperich_2021}. It then detects and excludes non-English documents using Langdetect~\cite{Danilak_M_langdetect}. Finally, it discards duplicates and trivially short privacy policies. Building on top of the Dragnet model~\cite{Peters_Matthew_dragnet}, the Scraper sanitizes HTML policies by removing irrelevant elements such as HTML tags, advertisement, and UI features (e.g., navigation bar). For now, \ourtool{} considers privacy policies in HTML format only, though we could have easily extended the preprocessing to handle less common formats such as PDF or raw text. The process results in a total of 51,781 usable privacy policies in plain text format.

\section{MTurker Selection and Survey}
\label{appendix_qual}

\textbf{MTurker Selection:} \ourtool{} selects mTurkers as survey participants based on requirements and tests. All participants should have an approval rate $>90$ and the number of HITs approved $>500$ while being an English speaker and an Android mobile user. The qualification test determines whether an mTurker has the capability to label privacy policies, in which mTurkers read through detailed annotation instructions and answer qualification questions. %
\ourtool{} also ensures no worker answers the same question repeatedly by recording and checking their worker IDs before assigning HITs.

\noindent\textbf{MTurk Survey:} \ourtool{} sends HITs\footnote{A \textbf{H}uman \textbf{I}ntelligence \textbf{T}ask is the term that the Amazon mTurk uses to describe a task.} as batches of multi-question surveys to Amazon mTurk. Each survey queries a worker for labels on data category relevance and action modes for each data action for a single text segment. The format of the survey has gone through several iterations of revision to enhance clarity. The version with the highest inter-annotator agreement among 5 mTurk annotators was eventually selected. To control labeling cost, we perform some calibration to identify the optimal payment for completing a HIT.  
We prepare sample HITs based on labeled text segments from the APP-350 and OPP-115 datasets to evaluate how worker wages affect the inter-annotator agreement rate and accuracy of the labels. We observe a significantly low agreement rate (31.9\%) for hourly payments below $\$1$\footnote{All amounts are in USD.}, while the agreement rate fails to increase significantly for payments higher than $\$5.2$ per hour (76.0\% for $\$5.2$ vs. 77.0\% for $\$13$). Similarly, the accuracy did not increase for payments above this threshold either.  Based on this, each HIT consists of roughly 40 surveys and the payment varied $\$0.16-0.25$ per survey. We also embed an honesty checker in each survey for mTurkers to indicate whether they paid close attention and answer questions with their best effort. MTurkers receive full payments regardless of what they answer but \ourtool{} discards all labels with ``No'' and re-publishes the same samples to be labeled by other qualified mTurkers. Manual inspection shows that 63\% of these are indeed incorrect, while the remaining false positives may still be random guesses that fall into the correct categories by chance. The conclusion aligns with the previous study ~\cite{rouse_reliability_2015}.

\begin{table} 
\centering
\footnotesize

\begin{tabular}{l|l|l|l|l}
~           & 50K  & 10K-50K & 1K-10K & 1K    \\ 
\hline
collect/use & 0.21 & 0.38    & 0.34   & 0.31  \\ 
\hline
share       & 0.21 & 0.31    & 0.38   & 0.28  \\ 
\hline
store       & 0.08 & 0.19    & 0.15   & 0.24  \\ 
\hline
overall average & 0.50 & 0.89    & 0.87   & 0.83 
\end{tabular} 
\caption{Average \# of explicit denials per privacy policy by \# of downloads }\label{tab:analysis-q1-downloads}
\end{table}

\begin{table} 
\centering
\footnotesize

\begin{tabular}{l|l|l|l|l|l}
~             & 4.5  & (4.0,
  4.5] & (3.0,
  4.0] & 3.0  & None  \\ 
\hline
collect/use   & 0.26 & 0.35         & 0.33         & 0.33 & 0.40  \\ 
\hline
share         & 0.32 & 0.40         & 0.39         & 0.24 & 0.32  \\ 
\hline
store         & 0.14 & 0.24         & 0.04         & 0.10 & 0.19  \\ 
\hline
overall
  average & 0.72 & 0.99         & 0.76         & 0.67 & 0.91 
\end{tabular} 
\caption{Average \# of explicit denials per privacy policy by rating}\label{tab:analysis-q1-rating}
\end{table}

\begin{table}
\centering
\footnotesize

\begin{tabular}{l|l|l|l}
                & assert & denial & choice  \\ 
\hline
collect/use     & 5.06   & 0.90   & 0.63          \\
share           & 1.41   & 1.07   & 0.89          \\
store           & 0.33   & 0.49   & 0.13          \\ 
\hline
overall
  freq. & 6.80   & 2.46   & 1.65         
\end{tabular} 
\caption{Average \# of labels per privacy policy by data actions and action modes} \label{tab:analysis-num-labels}
\end{table}

\begin{table*}
\small
\centering
\begin{tabular}{l|l|l|l|l|l|l|l|l} 
\diagbox[width=10em]{\footnotesize data category}{\footnotesize app category}
           & education & entertainment & music \& audio & tools & shopping & health \& fitness & business & average   \\ 
\hline
contact    & 0.93      & 0.92          & 0.77         & 0.63  & 1.00     & 1.00            & 0.57         & 0.82  \\
device     & 0.26      & 0.59          & 0.32         & 0.55  & 0.23     & 0.82            & 0.37         & 0.42  \\
location   & 0.16      & 0.30          & 0.30         & 0.52  & 0.85     & 0.47            & 0.12         & 0.38  \\
demographic     & 0.69      & 0.14          & 0.13         & 0.05  & 0.52     & 0.73            & 0.05         & 0.31  \\
health     & 0.03      & 0.02          & 0.00         & 0.03  & 0.00     & 1.13            & 0.22         & 0.18  \\
financial  & 0.31      & 0.65          & 0.65         & 0.18  & 0.93     & 0.32            & 0.08         & 0.40  \\
survey~    & 0.08      & 0.02          & 0.02         & 0.02  & 0.02     & 0.05            & 0.03         & 0.03  \\
personal identifier   & 0.13      & 0.08          & 0.02         & 0.00  & 0.08     & 0.03            & 0.02         & 0.05  \\
social media & 0.54      & 0.79          & 0.70         & 0.17  & 0.67     & 0.32            & 0.23         & 0.45 
\end{tabular}
\caption{Average number of text segments per app for each data category by Android app category }
\label{tab:analysis-datacate-appcate}
\end{table*}

\section{Sensitivity study} \label{evaluation_sensitivity}

\label{}
\begin{figure}
	\begin{minipage}{.5\linewidth}
		\begin{tabular}{@{}c@{}}
			\includegraphics[scale=0.38]{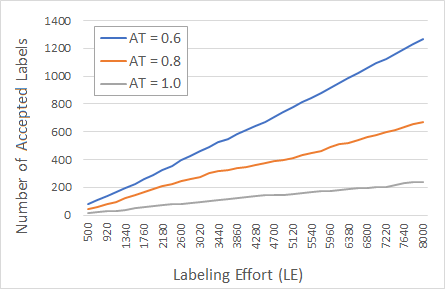} \\
			\footnotesize (a)  	
		\end{tabular} 
	\end{minipage}%
	\begin{minipage}{.5\linewidth}
		\begin{tabular}{@{}c@{}}
			\includegraphics[scale=0.38]{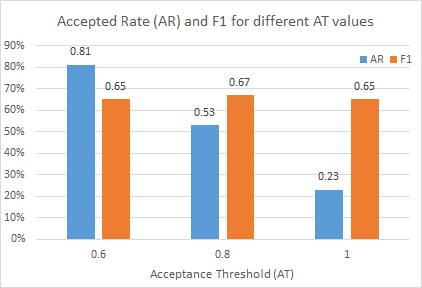} \\
			\footnotesize (b) 
		\end{tabular}
	\end{minipage}
	\caption{(a) an example result of Accepted labels vs. Labeling Requests at different Acceptance Thresholds (AT) (data category: contact); (b) average Acceptance Rate (AR) and F1 performance for different ATs }\label{fig:AT_eval}
\end{figure}

\textbf{Acceptance Threshold (AT)}: \ourtool{} uses an AT of $4/5$, which the inter-annotator agreement rate has to meet or exceed for the labels from the annotators to be accepted.  We investigate the effect different ATs have on the acceptance rate and accuracy of the resulting model.  We have the active learning select 1600 segments to be labeled, and have each of them labeled by 5 crowdsourced annotators for a total of 8000 labels.  We then apply 3 different ATs: 3/5, 4/5 and  5/5 and compute the number of accepted labels, average accept rate and resulting accuracy of the trained model. Figure~\ref{fig:AT_eval}-a %
shows the number of accepted labels by AT.  As expected, the lower the threshold, the larger the number of accepted labels, though we do observe that the variability in the number of accepted labels does increase with AT.  Similarly, Figure~\ref{fig:AT_eval}-b shows that the average acceptance rate also decreases inversely with AT.  However, we observe an interesting trend in the accuracy of the trained model, which peaks slightly at the 4/5 AT.  We surmise that the 3/5 threshold likely has lower accuracy due to a higher rate of incorrect labels due to the more permissive AT, while the 5/5 AT has too few training points to enable the model to generalize. Interestingly, combined with our results from Section~\ref{sec:eval:classification}, it shows that despite discarding some number of samples that don't meet the acceptance rate, \ourtool{} is still able to find enough representative samples to achieve higher classification accuracy than previous works that use all samples from trained annotators.

\noindent\textbf{Relabeling:} We evaluate the trade-off between Label \& Discard and Incremental Relabeling, as described in Section~\ref{subsec:design_relabel}.  We introduce a measure, the Relabeling Success Rate (RSR), which we define as the fraction of wasted labels that are recovered as accepted after relabeling.  We take 100 wasted labeling requests and republish them on mTurk to compute the RSR.  Out of the 100 relabeled segments, only 2 passed the AT after $N=2$ and only 1 passed the AT after $N=3$, resulting in an overall RSR of $3\%$ for 3 relabeling attempts.  This result heavily suggests that Incremental Relabeling is not an effective use of labeling resources.

This result is consistent with our earlier measurements, which suggest that the labels produced by an agreement rate of 4/5 across 5 mTurkers are reliable, as they also largely agree with labels produced by trained annotators.  It is not necessary, and does not help, to gather labels from a larger set of mTurkers.  Indeed, further investigation suggests that the wasted labeling requests may be on text segments that are inherently ambiguous. We also observe that the segments that end up wasting labeling resources tend to be longer and more complex, which contributes to a higher possibility of multiple interpretations. Specifically, we find that the average length of accepted segments is 380.4 characters while that of wasted segments is 439.2, a difference of roughly 15\%.  One of the wasted text segments is:
\begin{quoting}
\textit{``Health Plan may also partner with your Group Health Plan, if you obtain benefits through such a Plan, in which the Group Health Plan is responsible for some of the information being provided through the Health Plan site.''}    
\end{quoting} 
Both the grammar and meaning of this sentence are very ambiguous.  We find it notable that roughly 27\% of segments selected by \ourtool{} fall into this ambiguous category and believe investigating this is interesting future work.

\begin{figure}
	    \centering
	    \includegraphics[scale=0.65]{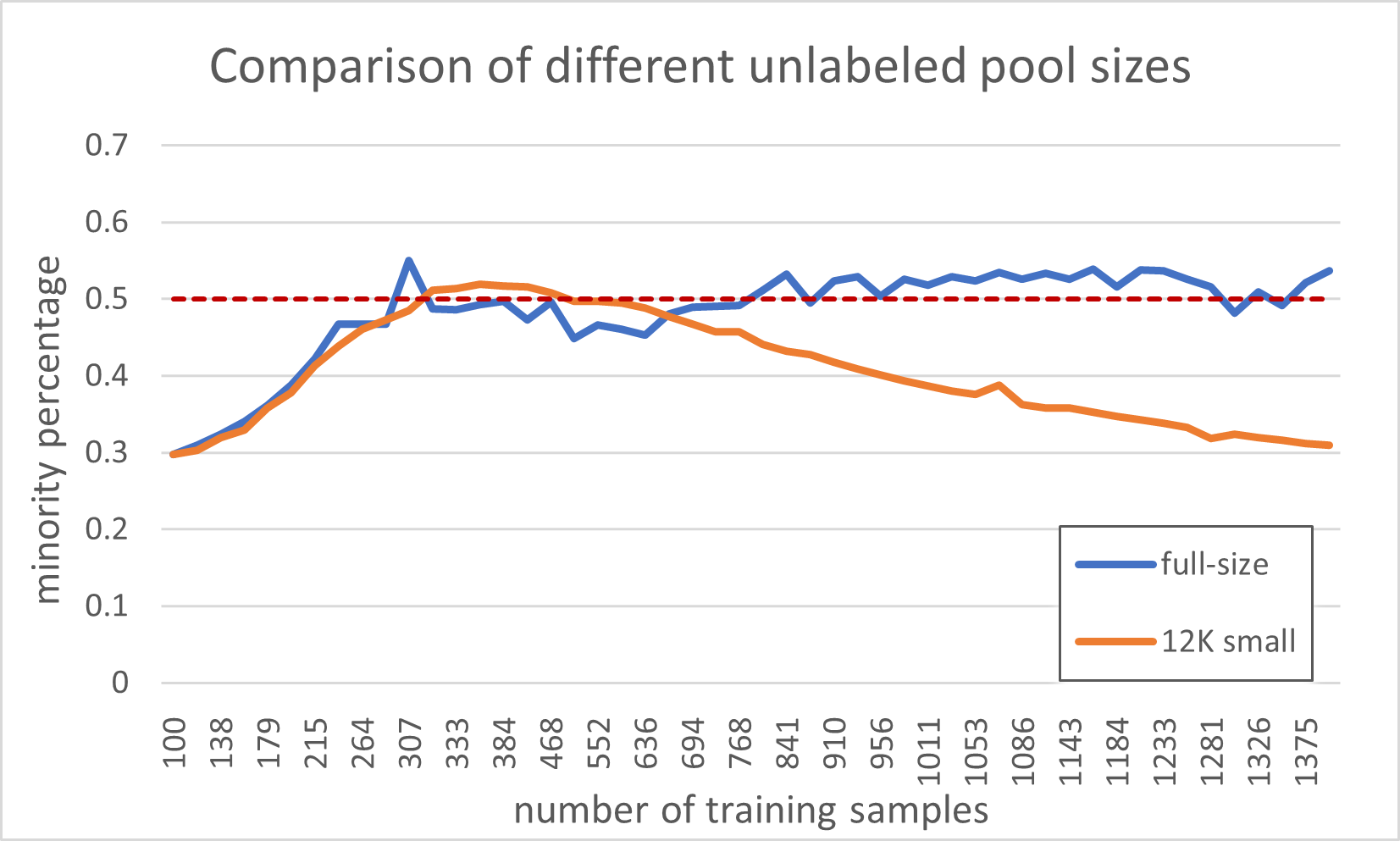}
	    \caption{Comparison between Contact Category Models trained on different sizes of the unlabeled training pool. The red dotted line shows the balance line of 50\% minority sample percentage}
	    \label{fig:balance_graph}
	\end{figure}

\noindent\textbf{Unlabeled pool:} We study how the size of an unlabeled training pool affects the training process. We experiment with \ourtool{} using two different pool sizes: one that contains the entire pool of all 3311K segments from 52K privacy policies and another that contains only 12K randomly selected segments. Here we focus on contact, device and location data categories for the \textit{collect/use} data action, which results in approximately 4K segments for each category. Otherwise, the two models share the exact same structure and hyperparameters. We train each model with an early stop when the number of training samples equals to 2000. We repeat each experiment five times and calculate the average accuracy. We observe a lower accuracy for all models trained using a smaller unlabeled pool: an average F1 of 86.7\% for models using the full-size unlabeled pool versus an average F1 of 90.2\% for models using the 12K unlabeled pool. Recall that the majority class is the assertion class (i.e. we are collecting) and the minority class is the denial class (i.e. we do not collect).  We found that the accuracy difference is mostly in the minority class, where the model with the smaller unlabeled pool is more likely to misclassify true minorities as false majorities. Figure~\ref{fig:balance_graph} compares the percentage of minority samples of the device data category as samples for the training set are selected.  In both the full and 12K small pool, we initially observe a fast increase in minority samples as the active learning selects the minority samples for labeling. However, for the smaller pool, the minority percentage soon starts to decrease as active learning exhausts all the minority samples in the pool and is forced to select those in the majority class.  As a result, the model does not generalize as well for the minority class, resulting in lower accuracy.  For the large pool, the makeup remains balanced at approximately half in both classes.  This is also confirmed by the final minority percentage as shown in Table~\ref{table:discriminator_eval}. All ending percentages of models trained with the full-size unlabeled pool converge to approximately 50\%, achieving a balance between the majority and minority classes.   

\section{Active Learning Querying Strategies} \label{appendix_querying}

\textbf{Basic Uncertainty} selects the least certain instance from the unlabeled set and requests for it to be labeled~\cite{Culotta_McCallum_2005}. The confidence probability is calculated using: 
\begin{equation}
	\underset{y \in Y}{max} [1- \mathbb{P}(y \in Y |\mathbf{x})]
\end{equation}     

\noindent\textbf{Margin Sampling} selects instances where the difference between the first most likely and second most likely classes are the smallest, minimizing the classification error~\cite{goos_active_2001}. \textbf{Entropy Sampling} chooses samples with the largest entropy in class probabilities, minimizing the log loss~\cite{shannon_mathematical_2001}. For binary classification, both are reduced to the basic uncertainty strategy. The three methods will query instances with a class posterior closest to 0.5, the most \textit{ambiguous} segments~\cite{settles_2012}. We confirmed this conclusion using our CPPS dataset. %

\section{Android App Analysis} \label{appendix_app-analysis}
Table~\ref{tab:analysis-q1-downloads},  Table~\ref{tab:analysis-q1-rating} and Table~\ref{tab:analysis-datacate-appcate} show the frequency of explicit denials by number of downloads, app rating, and app category, respectively. Table~\ref{tab:analysis-num-labels} summarizes the number of labels by data actions and action modes.

\end{document}